\documentclass[review]{elsarticle}

\usepackage{lineno,hyperref,subfigure,color}
\modulolinenumbers[1]

\usepackage{xspace} 
\def\Offline{\mbox{$\overline{\textrm%
      {Off}}$\hspace{.05em}\protect\raisebox{.4ex}%
    {$\protect\underline{\textrm{line}}$}}\xspace}%

\journal{Astroparticle Physics}

\begin{document}

\begin{frontmatter}

\title{Detection of ultra-high energy cosmic ray showers with a single-pixel fluorescence telescope}

\author[1,2]{T. Fujii\corref{mycorrespondingauthor}}
\ead{fujii@kicp.uchicago.edu}

\author[3]{M. Malacari}


\author[4]{M. Bertaina}
\author[5]{M. Casolino}
\author[3]{B. Dawson}
\author[9]{P. Horvath}
\author[9]{M. Hrabovsky}
\author[1]{J. Jiang}
\author[8]{D. Mandat}
\author[1]{A. Matalon}
\author[6]{J. N. Matthews}
\author[1]{P. Motloch}
\author[8]{M. Palatka}
\author[8]{M. Pech}
\author[1]{P. Privitera}
\author[8]{P. Schovanek}
\author[5]{Y. Takizawa}
\author[6]{S. B. Thomas}
\author[8]{P. Travnicek}
\author[7]{K. Yamazaki}


\address[1]{Kavli Institute for Cosmological Physics, University of Chicago, Chicago, IL, USA}
\address[2]{Institute for Cosmic Ray Research, University of Tokyo, Kashiwa, Chiba, Japan}
\address[3]{Department of Physics, University of Adelaide, Adelaide, S.A., Australia}
\address[4]{Dipartimento di Fisica, Universit\`a di Torino and INFN Torino, Torino, Italy}
\address[5]{RIKEN Advanced Science Institute, Wako, Saitama, Japan}
\address[9]{Palacky University, RCPTM, Olomouc, Czech Republic}
\address[8]{Institute of Physics of the Academy of Sciences of the Czech Republic, Prague, Czech Republic}
\address[6]{High Energy Astrophysics Institute and Department of Physics and Astronomy, University of Utah, Salt Lake City, Utah, USA}
\address[7]{Graduate School of Science, Osaka City University, Osaka, Osaka, Japan}

\cortext[mycorrespondingauthor]{Corresponding author}


\begin{abstract} 
We present a concept for large-area, low-cost detection of ultra-high energy cosmic rays (UHECRs) with a Fluorescence detector Array of Single-pixel Telescopes (FAST), addressing the requirements for the next generation of UHECR experiments. In the FAST design, a large field of view is covered by a few pixels at the focal plane of a mirror or Fresnel lens.
We report first results of a FAST prototype installed at the Telescope Array site, consisting of a single 200 mm photomultiplier tube at the focal plane of a 1~m$^2$ Fresnel lens system taken from the prototype of the JEM-EUSO experiment. The FAST prototype took data for 19 nights, demonstrating remarkable operational stability. We detected laser shots at distances of several kilometres as well as 16 highly significant UHECR shower candidates.
\end{abstract}


\end{frontmatter}


\section{Introduction}
\label{sec:introduction}
The origin and nature of ultra-high energy cosmic rays is one of the most intriguing mysteries in particle astrophysics~\cite{bib:Watson}. 
Given their minute flux, less than one per century per square kilometre at the highest energies, a very large area must be instrumented to collect significant statistics. 
The energy, arrival direction, and mass composition of UHECRs can be inferred from studies of the cascades of secondary particles (Extensive Air Shower, EAS) produced by their  interaction with the Earth's atmosphere. Two well-established techniques are used for UHECR detection: 1) arrays of detectors (e.g.  plastic scintillators, water-Cherenkov stations) sample EAS particles reaching the ground; 2) large-field-of-view telescopes allow for reconstruction of the shower development in the atmosphere by imaging  UV fluorescence light from atmospheric nitrogen excited by EAS particles. 


The Pierre Auger Observatory (Auger)~\cite{bib:auger}~\cite{bib:augerFD}, the largest UHECR experiment in operation, combines the two techniques, with arrays of particle detectors overlooked by fluorescence detector (FD) telescopes.
Auger covers an area of over 3,000 km$^2$ close to the town of Malarg\"{u}e in the province of Mendoza, Argentina.
The Telescope Array experiment (TA)~\cite{bib:tafd}~\cite{bib:tasd} is the second largest experiment in operation and uses the same detection techniques as Auger. 
TA is located near the town of Delta in central Utah, USA, and covers an area of 700~km$^2$. 
The High Resolution Fly's Eye experiment (HiRes)~\cite{bib:hires}, which consisted solely of FD stations (HiRes-I and HiRes-II), was operated from 1998 to 2006 on the U.S. Army's Dugway Proving Ground in western Utah.

Significant advances in our understanding of UHECRs have been achieved in the last decade by these experiments~\cite{bib:uhecr_review}. The existence of a  strong suppression of the cosmic ray flux above $10^{19.7}$ eV
is now unequivocally established~\cite{bib:spectrum_auger}~\cite{bib:spectrum_ta}. This observation is consistent with UHECRs being attenuated by interaction with the cosmic microwave background over distances of $\sim 100$~Mpc, as predicted by Greisen, Zatsepin and Kuzmin (GZK)~\cite{bib:gzk1}~\cite{bib:gzk2} in 1966. However, a cutoff in the spectrum of UHECRs at the accelerating sources may also offer an explanation.    
The mass composition reported by Auger through $X_{\max}$ (the depth in the atmosphere at which the EAS reaches its maximum energy deposit) suggests a transition from light nuclei at around $10^{18.3}$ eV to heavier nuclei up to energies of $10^{19.6}$ eV~\cite{bib:mass_auger}~\cite{bib:mass_implication_auger}. 
The mass composition reported by TA and HiRes is lighter, and consistent with a protonic composition for cosmic rays with energies greater than 10$^{18.2}$ eV~\cite{bib:hires_composition}~\cite{bib:ta_composition}.
However, mean $X_{\max}$ values observed by both Auger and TA are compatible within statistical uncertainties~\cite{bib:composition_wg}~\cite{bib:composition_wg2014}.
No evidence for photons or neutrinos in the UHECRs has been found thus far~\cite{bib:photon_auger}~\cite{bib:photon_ta}~\cite{bib:neutrino_auger}. Arrival directions of UHECRs are found to correlate with nearby extragalactic objects at a modest  2-3$\sigma$ significance level~\cite{bib:anisotropy_auger}~\cite{bib:anisotropy_ta}. Recently, TA has reported evidence of a hotspot in the northern hemisphere with a 3.4$\sigma$ post-trial significance~\cite{bib:hotspot_ta}.

These results are limited by statistics at the highest energies due to the GZK-like suppression. To further advance the field, the next generation of experiments will require an aperture which is larger by an order of magnitude. This may be accomplished by the fluorescence detection of UHECR showers from space, as in the proposed JEM-EUSO~\cite{bib:jemeuso} mission, or with a ground array much larger than Auger. Low-cost, easily-deployable detectors will be essential for a ground-based experiment. 

In this paper, we present an FD telescope concept which would fulfill these requirements, while maintaining adequate energy, $X_{\max}$ and angular resolution.  The  Fluorescence detector Array of Single-pixel Telescopes (FAST) would consist of compact FD telescopes featuring a smaller light collecting area and many fewer pixels than current generation FD designs, leading to a significant reduction in cost. The FAST design may be an attractive option not only for future UHECR experiments, but also for upgrades of existing UHECR observatories. For example, it could provide low-cost fluorescence detector coverage to the fourfold expansion of the TA experiment \cite{bib:TAx4}, and be used at the Pierre Auger Observatory to increase the number of showers detected in stereo with more than one telescope.
We present the FAST concept and its expected performance from simulations in Section~\ref{sec:fast}.
The FAST prototype installed at the TA site is described in Section~\ref{sec:fastproto}, with details of its operation and calibration given in Section~\ref{sec:test_measurements}.
Detection of UHECR showers with the FAST prototype is reported in Section~\ref{sec:shower_search}. 
Finally, conclusions are drawn in Section~\ref{sec:conclusions}.

\section{FAST concept and expected performance}
\label{sec:fast}
In the current Auger FD telescope design, a mirror system (effective light collecting area $A \sim 3$~m$^2$) reflects a $\sim 30^\circ \times 30^\circ$ patch of the sky onto a focal plane composed of 440 40 mm photomultiplier tubes (PMTs)~\cite{bib:augerFD}. In the FAST design, the same field of view (FOV) is covered by just a few $\sim200$ mm PMTs at the focal plane of a mirror or Fresnel lens of $A \sim 1$~m$^2$. We expect a significant cost reduction thanks to FAST's compact design with smaller light collecting optics, a smaller telescope housing, and a small number of PMTs and associated electronics. We estimate that the FAST reference design - a telescope of 1~m$^2$ effective area with a $\sim 30^\circ \times 30^\circ$ camera consisting of four PMTs - could cost less than 10\% of a current generation FD telescope with the same FOV coverage. In this work, we focus our simulation and experimental efforts on the reference design, as it is the most cost effective for detection of the highest energy showers ($>10^{19.5}$~eV). Increasing the number of PMTs would result in higher costs, with the primary gain being a lower energy threshold and improved efficiency and resolution at lower energies where large statistical samples have already been collected by current generation FDs. FAST stations, powered by solar panels and with wireless connection, could be deployed in an array configuration to cover a very large area. 

The proposed FAST design differs notably in operation from present generation FDs in two ways. The average current produced by the the night-sky background (NSB) in a fluorescence detector PMT is proportional to  $A \Delta \Omega$, where $\Delta \Omega$ is the pixel solid angle. In our design, $A_{\rm{FAST}} \sim 1$~m$^2$ and the pixel opening angle is $\sim 15^\circ$, compared with typical values of $\sim 3$~m$^2$ and $\sim 1.5^\circ$ for the Auger FD, or $\sim 7$~m$^2$ and $\sim 1^{\circ}$ for the TA FD. This means that FAST pixels will operate under significantly higher current. In addition, the relatively small detector aperture and large pixel solid angle leads to an increased energy threshold for UHECR detection, as the signal to noise ratio is proportional to $\sqrt{A/\Delta \Omega}$. The second major difference relates to shower geometry reconstruction. As current generation FD cameras consist of several hundred PMTs each viewing a small portion of the sky, a detected shower is seen as a line of triggered pixels which define what is known as the shower-detector plane. The shower orientation within this plane can then be determined either from the pixel timing information via a $\chi^2$ minimization, or via the intersection of two or more shower-detector planes if the same shower is seen by more than one FD station. As the proposed FAST design consists of only a few pixels, each viewing a large portion of the sky, geometry reconstruction of the shower detector plane in this manner is not possible. FAST would therefore need to be operated alongside a surface detector which independently provides the EAS geometry. Alternatively, a FAST-only reconstruction may be possible by combining measurements from several FAST stations. For a shower that triggers multiple FAST stations, the measured signals can be used to constrain the shower geometry. In this case, a ground array of water-Cherenkov tanks or scintillators would not be required, further reducing the cost. 
The potential of this geometry reconstruction method is currently being investigated, and will be reported elsewhere.
We foresee a trigger and data acquisition system similar to that of current generation SD arrays: a first level trigger operating at a rate of 20-100 Hz optimized for microsecond long pulses will be implemented in each FAST PMT, with the corresponding digitized data kept locally in a circular buffer of several seconds depth. FAST stations will send their trigger information to a central location which will reply with a readout request when appropriate conditions are met (e.g. the timing from different FAST stations is consistent with a shower candidate).

We have performed extensive simulations to study the performance of the FAST design with a geometry reconstructed by a surface array. Our reference telescope has an effective area of 1~m$^2$ and a $30^{\circ}\times30^{\circ}$ FOV camera consisting of four PMTs. A FAST station consists of twelve such telescopes, covering $360^\circ$ in azimuth. We present simulations for a triangular arrangement of FAST stations with a spacing of 20 km. UHECR showers were generated with CORSIKA \cite{bib:corsika}, and a modified version of the Auger \Offline software \cite{bib:auger_fdanalysis} was used for the FAST telescope simulation and shower reconstruction. Since the design of the telescope is not yet finalized, a generic telescope was simulated, which included an effective light collecting area of 1~m$^2$, a mirror, a UV band-pass filter, and four PMTs. The wavelength dependence of the mirror reflectivity, the UV filter transmission, and the PMT quantum efficiency were included in the simulation.   Simulated showers were thrown following a realistic zenith angle distribution, and cores were placed randomly within a circle of radius 10 km in the centre of the triangular arrangement at an altitude of 1400 m above sea level. 
Emission of fluorescence photons in proportion to the shower energy deposit along its path in the atmosphere was simulated according to precise laboratory measurements of the fluorescence yield~\cite{bib:airfly1}~\cite{bib:airfly2}.
Light attenuation in the atmosphere due to Rayleigh and Mie scattering was included in the simulation. Lastly, fluctuations in the PMT signal due to the night-sky background were included. Atmospheric attenuation parameters and a NSB level typical of the Auger and TA sites were assumed. 

\begin{figure}
    \includegraphics[width=1.0\linewidth]{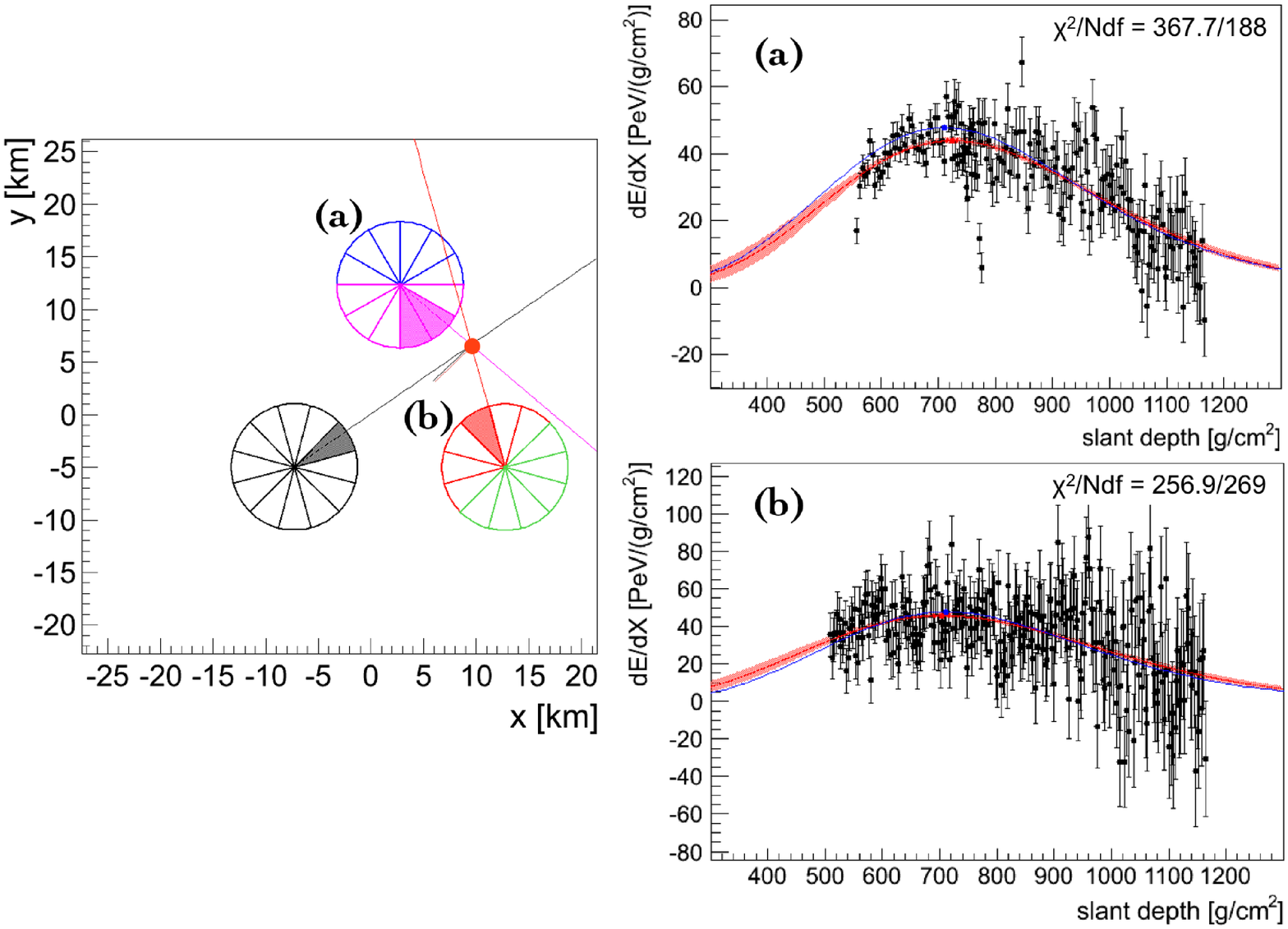}
    \caption{FAST reconstruction of a simulated 10$^{19.5}$ eV shower. Left panel: FAST station layout, with the shower core location indicated by the red dot. Right panels: reconstructed shower energy deposit profiles using a geometry smearing of 1.0$^{\circ}$ in arrival direction and 100~m in core location (to simulate the geometry resolution of a surface detector array). Blue (red) lines indicate the simulated (reconstructed) shower parameters. }
    \label{fig:fast_sim}
\end{figure}

\begin{figure}
    \includegraphics[width=1.0\linewidth]{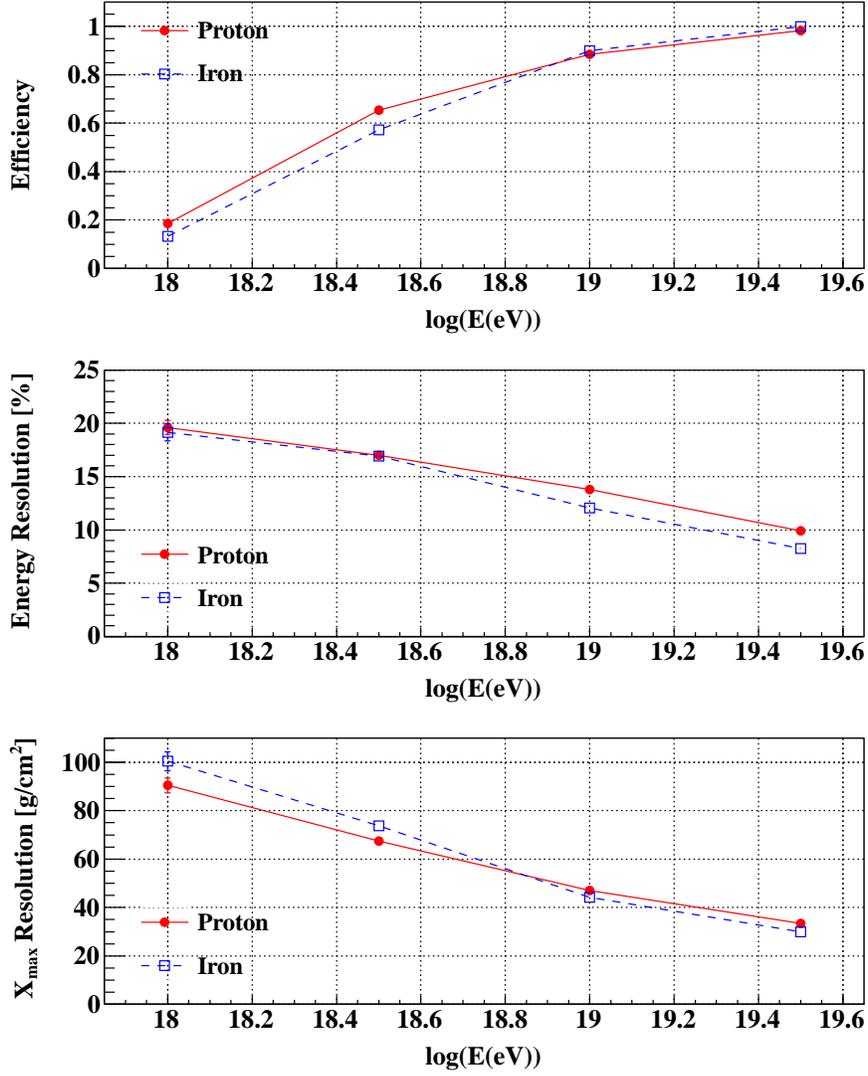}
    \caption{From top to bottom: FAST reconstruction efficiency, energy resolution and $X_{\max}$ resolution as a function of energy. Results are given for proton (red) and iron (blue) simulated showers. The FAST shower profle is reconstructed with a geometry smearing of 1.0$^{\circ}$ in arrival direction and 100 m in core location (to simulate the geometry resolution of a surface detector array).} 
    \label{fig:res_eff}
\end{figure}

An example of a simulated FAST event is shown in Figure~\ref{fig:fast_sim}. 
To reconstruct shower parameters from the FAST signal, the shower geometry must be given by a surface array. For the sake of simplicity, we did not simulate such an array, but rather took the true geometry of the simulated shower and smeared it by 1.0$^{\circ}$ in arrival direction and 100~m in core location, which are typical resolutions of existing UHECR surface arrays (e.g. \cite{bib:augerSD_resolution}). Given the geometry, the energy deposit profile at the shower axis is reconstructed by unfolding the detector efficiency and the atmospheric attenuation. The energy and $X_{\rm{max}}$ of the shower is then obtained from a Gaisser-Hillas~\cite{bib:gh_function} fit to the shower profile. To estimate the efficiency and resolution of the reference design we selected events with $X_{\max}$ reconstructed within the FOV, and with an observed slant depth greater than 200~g/cm$^2$. The reduced $\chi^2$ of the longitudinal profile fit was required to be less than 5. These are rather general selection criteria, which provide a sample of well reconstructed showers. The corresponding efficiency, energy resolution, and $X_{\max}$ resolution as a function of energy are given in Figure~\ref{fig:res_eff}.
As expected, FAST performs best at the highest energies ($>10^{19.5}$ eV), where reconstruction efficiency is close to 100\%, energy resolution is $\sim 10$\%, and $X_{\max}$ resolution is $\sim 34$ g/cm$^2$. The quality of the reconstruction at the highest energies is comparable to that of current generation FDs, making FAST a viable low-cost option for next generation UHECR experiments. 
 

\section{FAST prototype at the Telescope Array site}
\label{sec:fastproto}
A first test of the FAST concept was performed profiting from the existing infrastructure of the JEM-EUSO experiment at the TA site in Utah, USA, where a prototype~\cite{bib:euso_optics} is currently installed for a comprehensive test of the optics and electronics of this space-based detector. The light collecting area ($\sim1~\rm{m}^2$) and circular FOV ($\sim 7^{\circ}$ radius) of the JEM-EUSO prototype telescope (EUSO-TA telescope) are close to the FAST reference design (for a single pixel) providing a perfect test bed for the FAST concept. 

\begin{figure}
    \centering
    \subfigure[EUSO-TA telescope]{\includegraphics[width=0.7\linewidth]{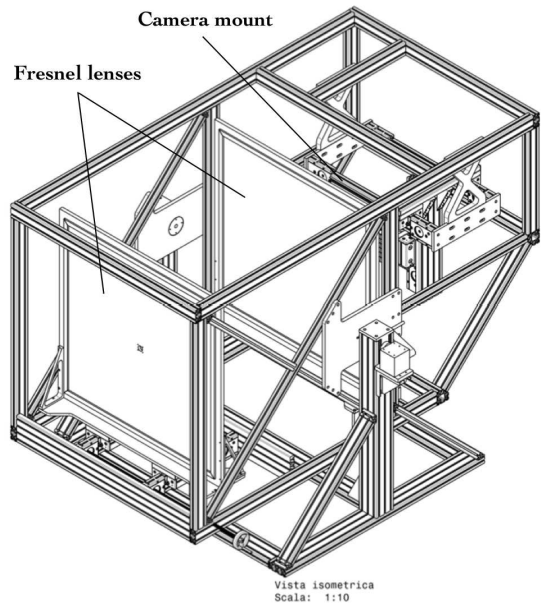}}
    \subfigure[FAST prototype camera]{\includegraphics[width=0.6\linewidth]{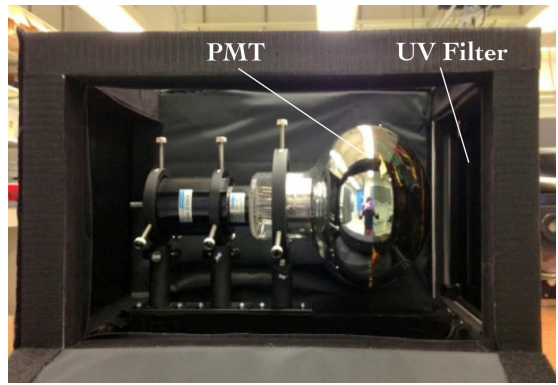}}
    \caption{The EUSO-TA telescope installed at the TA site, Utah, USA (a). The size of the telescope is approximately 1.8 m $\times$  2.0 m $\times$  2.6 m (H $\times$ W $\times$ L) . The FAST prototype camera, consisting of a single 200 mm PMT and a UV transparent filter, was installed at the focal plane of the telescope (b). The size of the camera is 0.34 m $\times$ 0.34 m $\times$ 0.48 m (H $\times$ W $\times$ L) .} 
    \label{fig:jemeuso_optics}
\end{figure}

The EUSO-TA telescope is hosted in a small hut in front of the TA FD building at the Black Rock Mesa site. Its optics consists of two $1~\rm{m}^2$ Fresnel lenses, with a UV transparent acrylic plate placed at the diaphragm for protection (Figure~\ref{fig:jemeuso_optics}). The telescope is exposed to the night sky by a manually operated shutter. For the purpose of the FAST test we installed a 200 mm PMT (R5912-03, Hamamatsu) at the focal plane of the telescope. A UV band-pass filter (Schott MUG-6 glass) was placed in front of the PMT to reduce the night sky background (NSB). The PMT was equipped with an AC-coupled active base (E7694-01, Hamamatsu) to maintain stable gain under the high current expected during operation. To track the PMT response, we attached to the PMT's surface a YAP pulsed light source consisting of a YAlO$_3$:Ce scintillator crystal excited by a 50 Bq $^{241}$Am source.

The electronics and data acquisition system (DAQ) of the FAST prototype were built from commercial modules. 
The PMT was kept at  +908 V high voltage (mod. N1470, CAEN), corresponding to a gain of $\sim 5 \times 10^4$. PMT signals from both the anode and last dynode were amplified ($\times 50$ by mod. 777, Phillips Scientific and $\times 10$ by mod. R979, CAEN, respectively) to provide a large dynamic range, and then passed through a low-pass filter before digitization by a 12-bit FADC (mod. SIS3350, Struck Innovative Systeme). The time duration of fluorescence signals varies from a few hundred nanoseconds for close-by showers to tens of microseconds for distant showers. A sampling rate of 10~MHz is found to optimize the signal-to-noise ratio for most showers, and has been adopted by the Auger and TA FDs. We use a 50~MHz sampling rate to minimize saturation in any single time bin, which is particularly effective for showers directed towards the telescope where the signal is compressed in time. Adjacent time bins are then summed to obtain an equivalent sampling rate of 10~MHz. 
The digitizer was hosted in a portable VME crate (mod. VME8004B, CAEN), together with a controller (mod. V7865, GE Intelligent platforms) and a GPS unit (mod. GPS2092, Hytec) providing event time stamps. 
Whenever any of the fluorescence telescopes in the adjacent TA building were triggered by a candidate UHECR shower an external trigger was issued to the FAST DAQ, with a typical rate of $\sim3$~Hz~\cite{bib:tafd_elec}.
In addition, a high-threshold internal trigger was periodically activated during data taking to collect YAP signals for monitoring purposes. The FAST DAQ was remotely controlled via a wireless network. 

\section{Measurements at the TA site}
\label{sec:test_measurements}
The FAST prototype operated for 19 days in April and June 2014 during clear, moonless nights for a total of 83 hours. 
Several measurements were performed to validate the FAST concept, including studies of the night-sky background, of the stability of the YAP signal, and of distant UV laser shots. In this section we outline the results of these measurements.

\subsection{Night sky background and stability}
\label{subsec:nsb}
The average current of a pixel in any FD telescope is dominated by the night sky background, typically $\sim$ 100 photons/deg$^2$/m$^2$/$\mu$s. Current generation FD telescopes have a small FOV when compared to the $\sim7^{\circ}$ radius of the FAST prototype.  Since the average pixel current is proportional to the light collecting area of the telescope and the pixel solid angle, we expect a significantly larger current in the FAST prototype. Stability of the PMT gain under these conditions must be verified. On the other hand, the FAST pixel should be quite insensitive to bright UV stars that would produce a large fractional increase in the current when entering a small FOV pixel. Also, the night-sky background must be well characterized since its fluctuations ultimately determine the energy threshold for UHECR detection. 

\begin{figure}
    \includegraphics[width=1.0\linewidth]{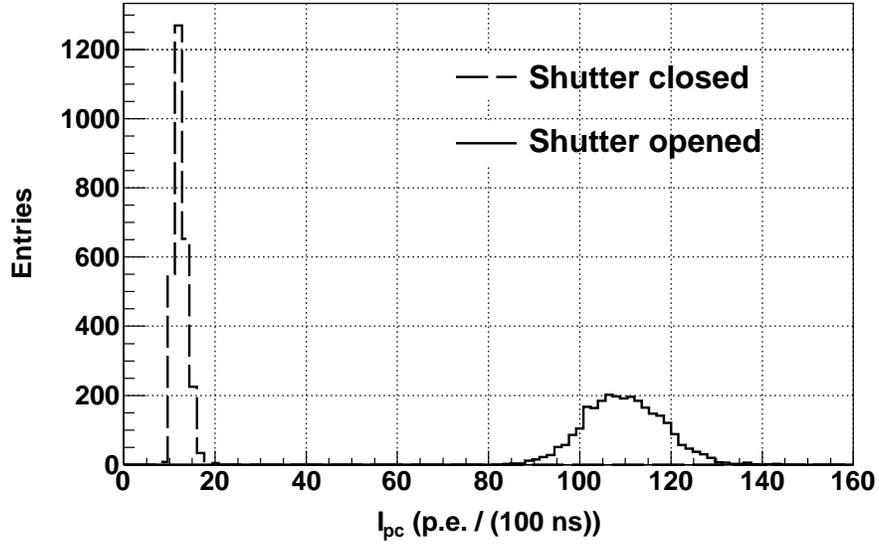}
    \caption{Photocathode current measured by the FAST PMT with the shutter closed (dashed line) and opened (solid line).}
    \label{fig:nsb}
\end{figure}

\begin{figure}
    \includegraphics[width=1.0\linewidth]{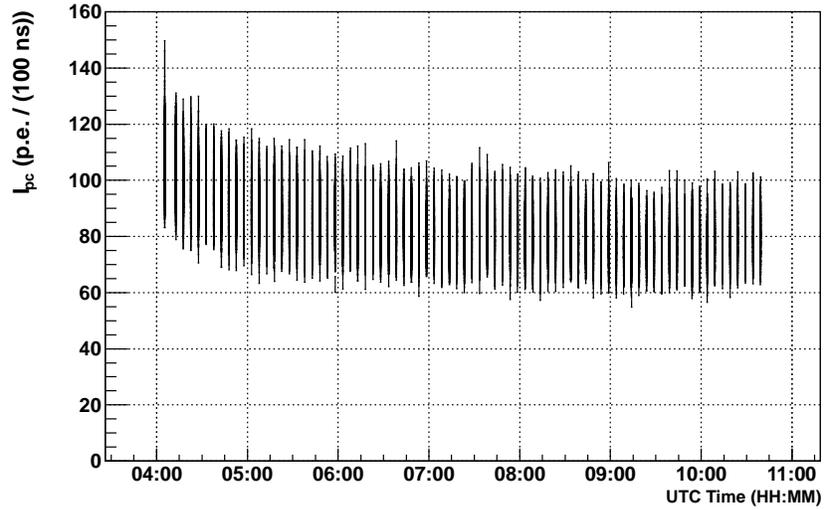}
    \caption{Stability of the photocathode current during a seven hour data taking run.}
    \label{fig:bg_stability}
\end{figure}

AC coupling of the FAST PMT does not allow for a direct measurement of the average current. However, fluctuations in the NSB are recorded as fluctuations in the PMT pedestal, whose variance $\sigma^2_{\rm{ADC}}$ is linearly related to the average photocathode current $I_{\rm{pc}}$ for a noiseless system~\cite{bib:gemmeke}. The variance of the FAST PMT had been previously calibrated in terms of photocathode current with a light source of known flux in a dedicated laboratory measurement.
We used this calibration to interpret the variance measured during field operation of the FAST prototype. A large increase in the FAST PMT variance is observed after opening the shutter at the beginning of the night (Figure~\ref{fig:nsb}). With the shutter closed, the variance is dominated by the noise in the FAST electronics chain, equivalent to $I_{\rm{pc}}=15$~p.e./100 ns. With the shutter opened, a photocathode current of 115 p.e./100 ns is measured, indicating that the electronic noise is negligible with respect to the NSB. The measured photocathode current is in good agreement with expectations. The NSB level detected by the TA FD ($\sim$ 100 photons/deg$^2$/m$^2$/$\mu$s) corresponds to a FAST current of $\sim$120~p.e/100~ns, estimated assuming a 7 degree circular FOV, a 20\% PMT quantum efficiency, and an average optical efficiency of 40\%. The r.m.s. fluctuations in the NSB, $\sigma_{\rm{NSB}} \sim$11 p.e./100 ns, dictate the sensitivity of the FAST prototype. 
The evolution of $I_{\rm{pc}}$ during seven hours of continuous data taking is shown in Figure~\ref{fig:bg_stability}. A smooth decrease as a function of time is observed, representing the change in the NSB during operation. We did not observe sudden jumps in the current, confirming that bright UV stars passing through the FAST FOV have a negligible effect.

\begin{figure}
    \includegraphics[width=1.0\linewidth]{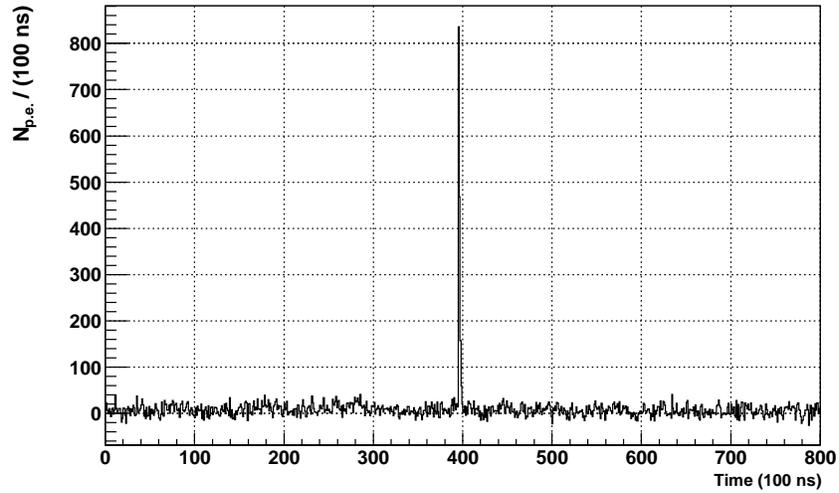}
    \caption{FADC signal recorded for a YAP light pulse. It is used to monitor the relative gain of the PMT.}
    \label{fig:yap}
\end{figure}

\begin{figure}
    \includegraphics[width=1.0\linewidth]{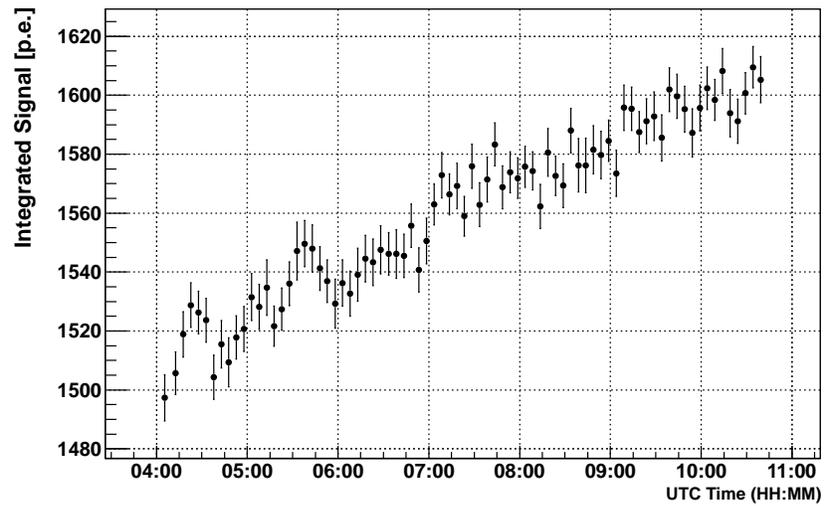}
    \caption{Variation of the YAP signal during a seven hour data taking run.}
    \label{fig:yap_stability}
\end{figure}

The FAST PMT gain was monitored during data-taking with the stable light pulses provided by the YAP source. An example of a digitized YAP signal is given in Figure~\ref{fig:yap}. 
The signal is given in units of photoelectrons per 100 ns and is obtained by summing 5 consecutive time bins at the nominal 50 Hz sampling rate (see Section~\ref{sec:fastproto}).
The measured variation in the YAP signal during a night is shown in Figure~\ref{fig:yap_stability}. The overall change is small ($\sim7\%$) and consistent with the known temperature dependence of the PMT gain ($\sim -1$\%$/^{\circ}$C). We expected an increase in the gain as the temperature drops during the night, since the housing of the FAST prototype is not temperature controlled.

\subsection{Detection of distant laser shots}
\label{subsec:laser}
UV laser shots are routinely used for calibration of FD telescopes and atmospheric monitoring~\cite{bib:clfauger}~\cite{bib:clf}. While traversing the atmosphere, the laser light side-scatters on air molecules and aerosol particles into the FD field of view, producing signals similar to a UHECR shower. The TA site is equipped with a Central Laser Facility (CLF), located about 21 km from the Black Rock Mesa site. It consists of a 355 nm UV laser which fires 300 vertical shots every 30 minutes during data taking. In addition, a Portable UV Laser System (PLS)~\cite{bib:portable_laser} can be deployed at different locations in the TA site. Both systems provide laser pulses of 2.2 mJ energy, approximately equivalent in intensity to a $\sim 10^{19.2}$ eV shower. We made extensive use of these laser facilities to characterize the performance of the FAST prototype. 

\begin{figure}
    \includegraphics[width=1.0\linewidth]{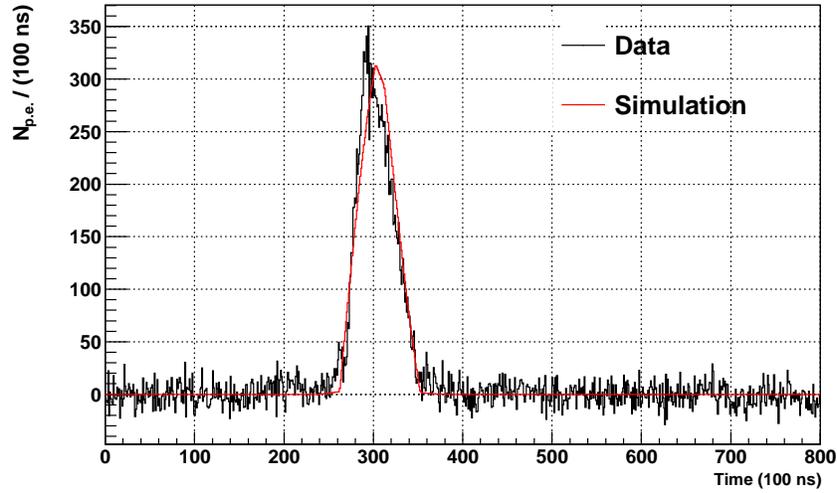}
    \caption{FADC signal corresponding to a vertical PLS laser shot at a distance of 6 km. The simulated signal is overlplotted in red and normalized to fit the measured peak.}
    \label{fig:portable_laser}
\end{figure}

\begin{figure}
    \includegraphics[width=1.0\linewidth]{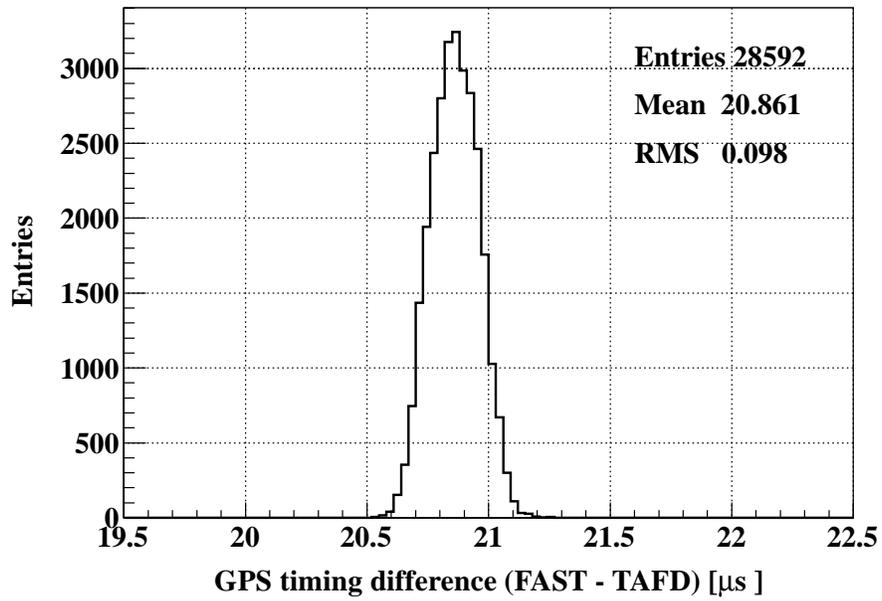}
    \caption{Difference between the TA FD and the FAST prototype GPS time for laser shots.}
    \label{fig:gps_timing}
\end{figure}

The signal measured by FAST for a single PLS shot is shown in Figure~\ref{fig:portable_laser}, with the PLS located at a distance of 6 km. The signal is well above the NSB level, and individual pulses were detected with 100\% efficiency. We used this data to calibrate the relative timing between FAST and the TA FD by comparing the GPS time recorded by the two detectors for the same laser shot.  An offset was expected, since the external trigger to the FAST DAQ required some processing time in the TA trigger board. The distribution of the difference between the FAST and TA fluorescence detector GPS times is shown in Figure~\ref{fig:gps_timing}. An offset of 20.86 $\mu$s was measured, attributed to the TA trigger processing time. The r.m.s. of $\sim100$~ns is consistent with the GPS resolution, and adequate for the purpose of the FAST prototype test. A precise measurement of this relative timing was essential in the search for UHECR showers presented in Section~\ref{sec:shower_search}.  

\begin{figure}
    \includegraphics[width=1.0\linewidth]{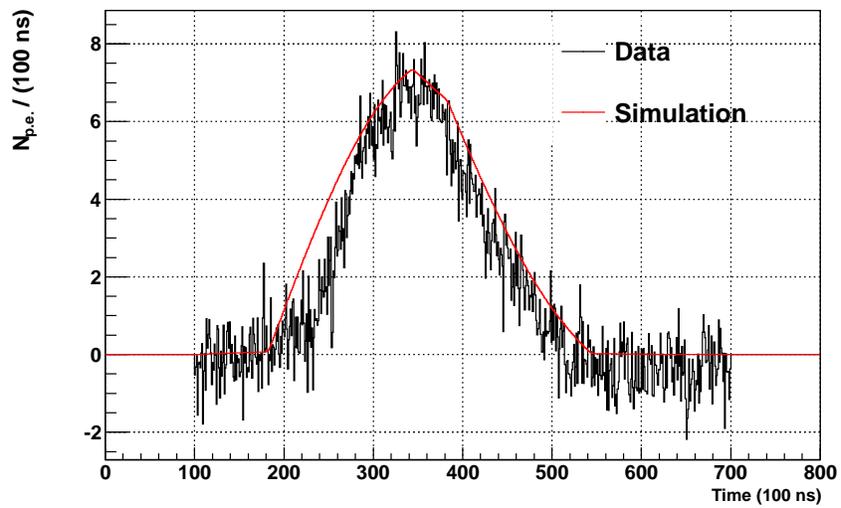}
    \caption{FADC signal corresponding to vertical CLF laser shots at a distance of 21 km. Since a single laser shot is at the limit of detection, 233 laser shots were averaged to improve the sensitivity. The red curve shows the expected signal from simulations of a 2.2 mJ vertical laser.} 
    \label{fig:clf}
\end{figure}

We also performed measurements with the CLF, whose laser shots passed right through the center of the FAST FOV. The CLF signal was expected to be attenuated to the limit of detectability due to its distance from the FAST prototype. Individual CLF laser pulses could not be resolved. However, a clear signal was observed when averaging over many laser shots (Figure~\ref{fig:clf}). The average signal amplitude was found to be 7 p.e./100 ns, indeed too small to allow for the detection of individual shots (compared with $\sigma_{\rm{NSB}}$, see Section \ref{subsec:nsb}). 

\begin{figure}
    \includegraphics[width=1.0\linewidth]{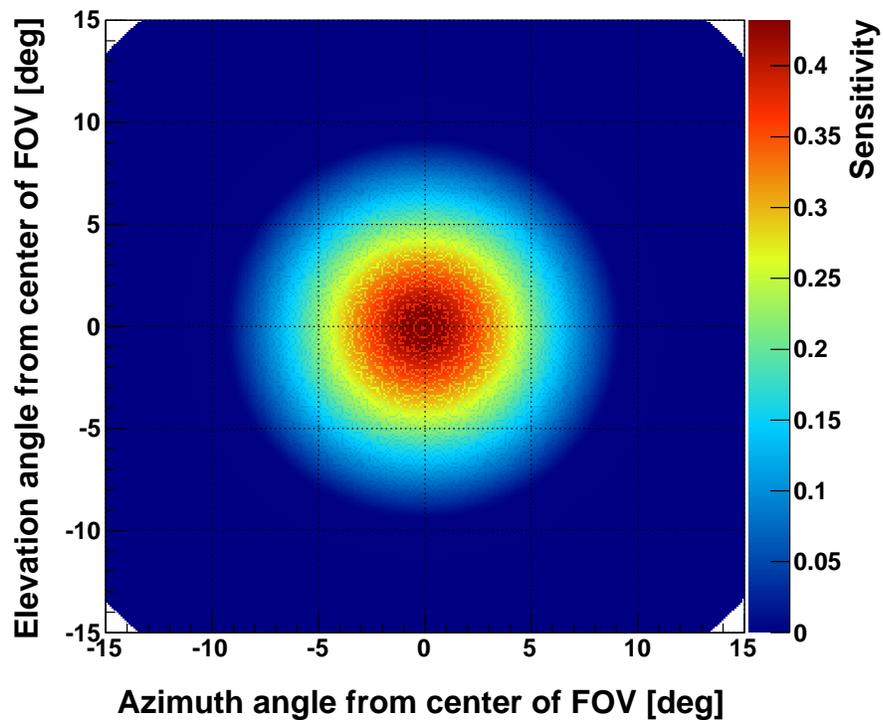}
    \caption{Efficiency of the FAST prototype's optics as a function of the angle to the optical axis, obtained with a ray tracing simulation of the telescope.}
   \label{fig:sensitivity}
\end{figure}

A simulation of the FAST prototype's response to laser shots was performed to compare with the PLS and CLF data. For this purpose, the efficiency of the EUSO-TA telescope as a function of angle was obtained from a ray-tracing simulation of the Fresnel lenses (Figure~\ref{fig:sensitivity}). This efficiency is defined as the ratio between the number of photons arriving at the camera to the number of photons injected at the aperture, and is calculated as a function of the injection angle relative to the optical axis. The FAST simulation includes the wavelength dependent quantum efficiency of the FAST PMT (measured in a dedicated laboratory setup before installation) and realistic light attenuation in the atmosphere due to Rayleigh and Mie scattering (with parameters typical of the TA site)~\cite{bib:lidar_ta}. The simulated signal is superimposed on the measured PLS and CLF laser shots in Figures~\ref{fig:portable_laser} and \ref{fig:clf}. As the laser shot energy of the PLS is not monitored, the simulated signal was rescaled by $-30\%$ to match the measured signal. The simulated CLF signal has not been rescaled as the laser energy is continuously monitored. 
Overall, there is good agreement between measurements and simulations, and any differences in the shape of the signal can be explained by the uncertainties in the optical efficiency (the ray tracing model assumes perfect Fresnel lenses), in the alignment of the EUSO-TA telescope and the FAST camera position, and in the assumed atmospheric attenuation. 

\section{Detection of UHECR showers}
\label{sec:shower_search}
Detection of very energetic showers ($> 10^{19}$ eV) in the limited running time of the FAST prototype was unlikely. However, we expected to observe a few lower energy close-by showers. A search was performed, driven by well reconstructed TA FD events which generated an external trigger for the FAST DAQ. 
 
\begin{figure}
    \subfigure[]{\includegraphics[width=1.0\linewidth]{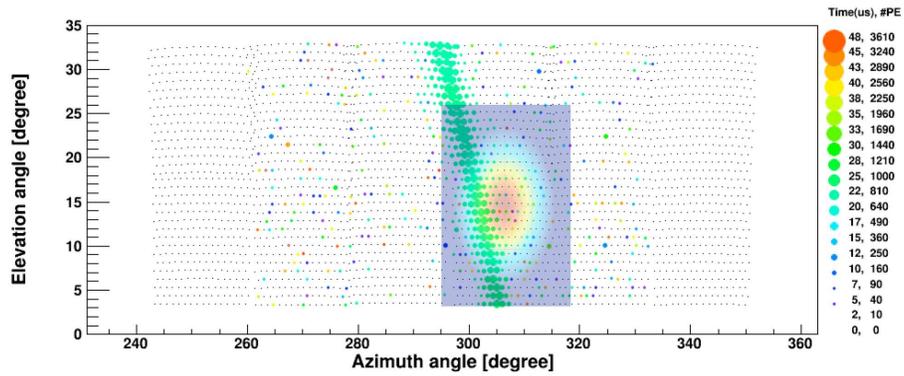}}
    \subfigure[]{\includegraphics[width=1.0\linewidth]{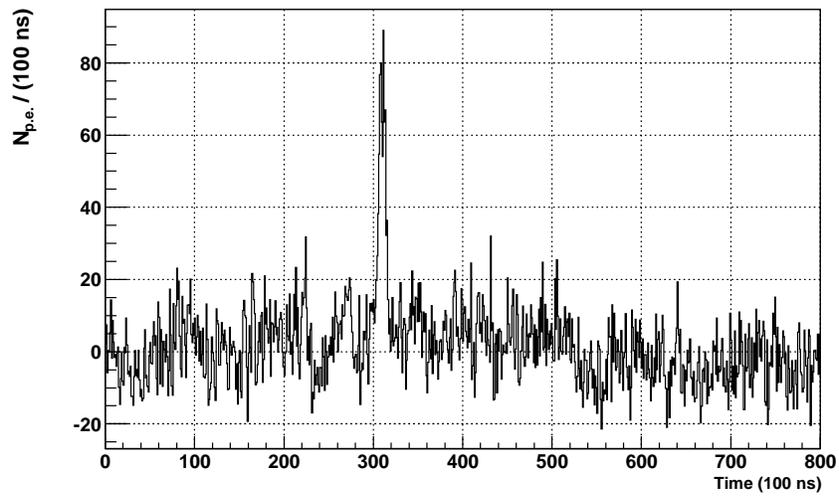}}
    \caption{A $10^{18}$ eV shower simultaneously detected by the TA FD and the FAST prototype. In (a), the shower is shown in the TA FD event display, with the FOV of the FAST prototype superimposed (see Figure~\ref{fig:sensitivity}). In (b), the corresponding FADC trace recorded by the FAST PMT.}
    \label{fig:shower}
\end{figure}

First we selected TA FD events with a reconstructed shower geometry passing through the FOV of the FAST prototype (Figure~\ref{fig:shower}.a). We then searched the corresponding FAST FADC traces for pulses with a maximum signal to noise ratio greater than 5$\sigma$, where $\sigma$ was calculated from the pedestal r.m.s. of the first 10~$\mu$s of the trace (Figure~\ref{fig:shower}.b). The search was performed in a time interval of 70~$\mu$s, positioned in the trace according to the relative timing between FAST and the TA FD (Section~\ref{subsec:laser}).  We found 16 shower candidates in the 83 hour dataset, with an estimated background of $<$ 1 event. The background was estimated from the data by applying the same search criteria to FAST traces recorded in coincidence with TA FD showers detected outside the FAST FOV. 

\begin{figure}
    \includegraphics[width=1.0\linewidth]{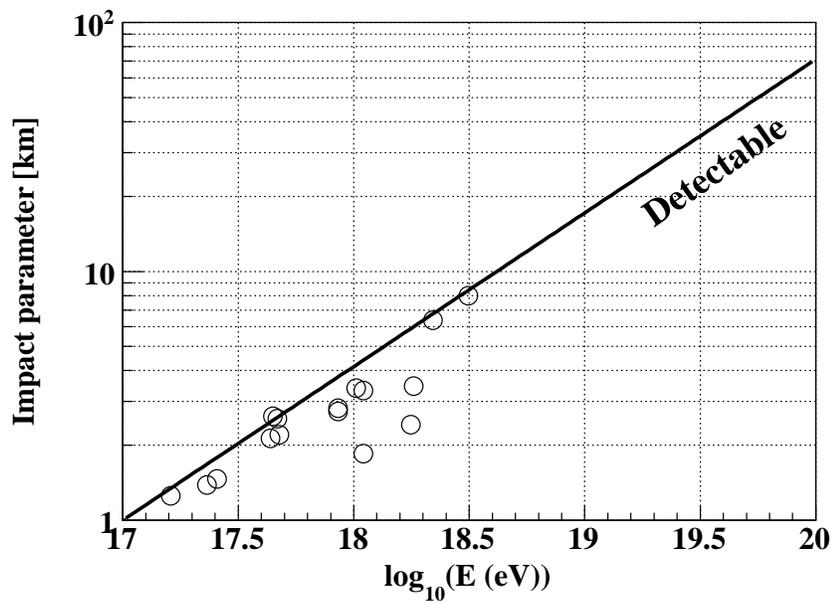}
    \caption{Correlation between the impact parameter and energy of the 16 cosmic ray shower candidates detected by the FAST prototype. Both shower parameters were obtained from the TA standard reconstruction. The line indicates the maximum detectable distance consistent with our limited data set.}
    \label{fig:shower_candidates}
\end{figure}

Although small, this sample provides an estimate of the sensitivity of the FAST prototype. The correlation between the impact parameter (i.e. the distance of closest approach of the shower axis to the FAST prototype) and the energy of the 16 showers is plotted in Figure~\ref{fig:shower_candidates}, with shower parameters given by the standard reconstruction of the TA FD~\cite{bib:tafd_reconstruction}. At any given energy, we expect showers to be detected up to a maximum impact parameter, $r_{\rm{det}}$. An approximate $r_{\rm{det}}$ bounding our limited data set is indicated by the line in Figure~\ref{fig:shower_candidates}.  When extrapolated to $10^{19.0}$ eV a maximum detectable distance of $\sim15$~km is obtained, which is a good indication of the validity of the concept introduced in Section~\ref{sec:fast}.

\begin{figure}
    \subfigure[Data: $E_{\rm{rec}}=10^{17.2}$ eV]{\includegraphics[width=0.49\linewidth]{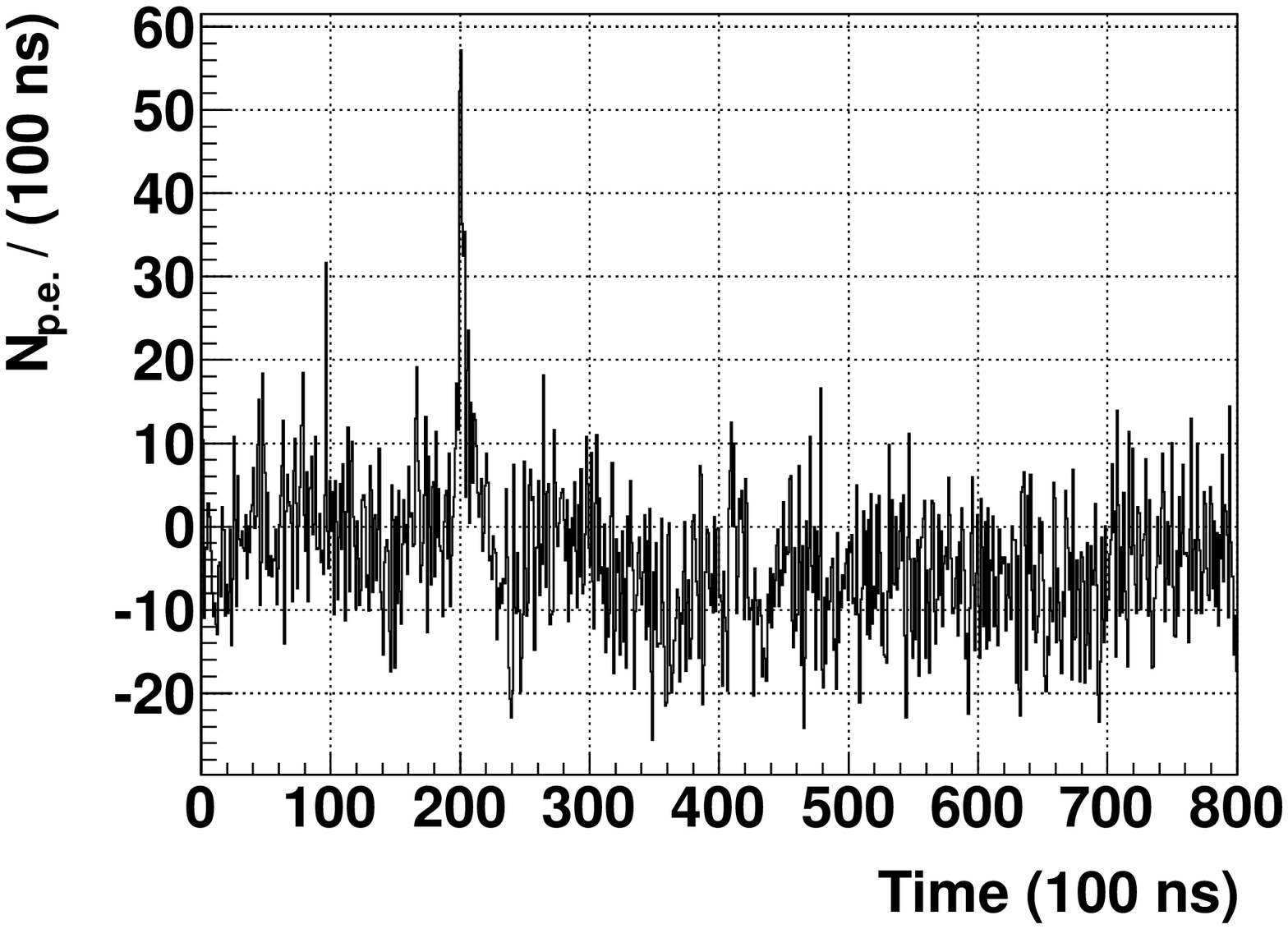}}
    \subfigure[Data: $E_{\rm{rec}}=10^{18.0}$ eV]{\includegraphics[width=0.49\linewidth]{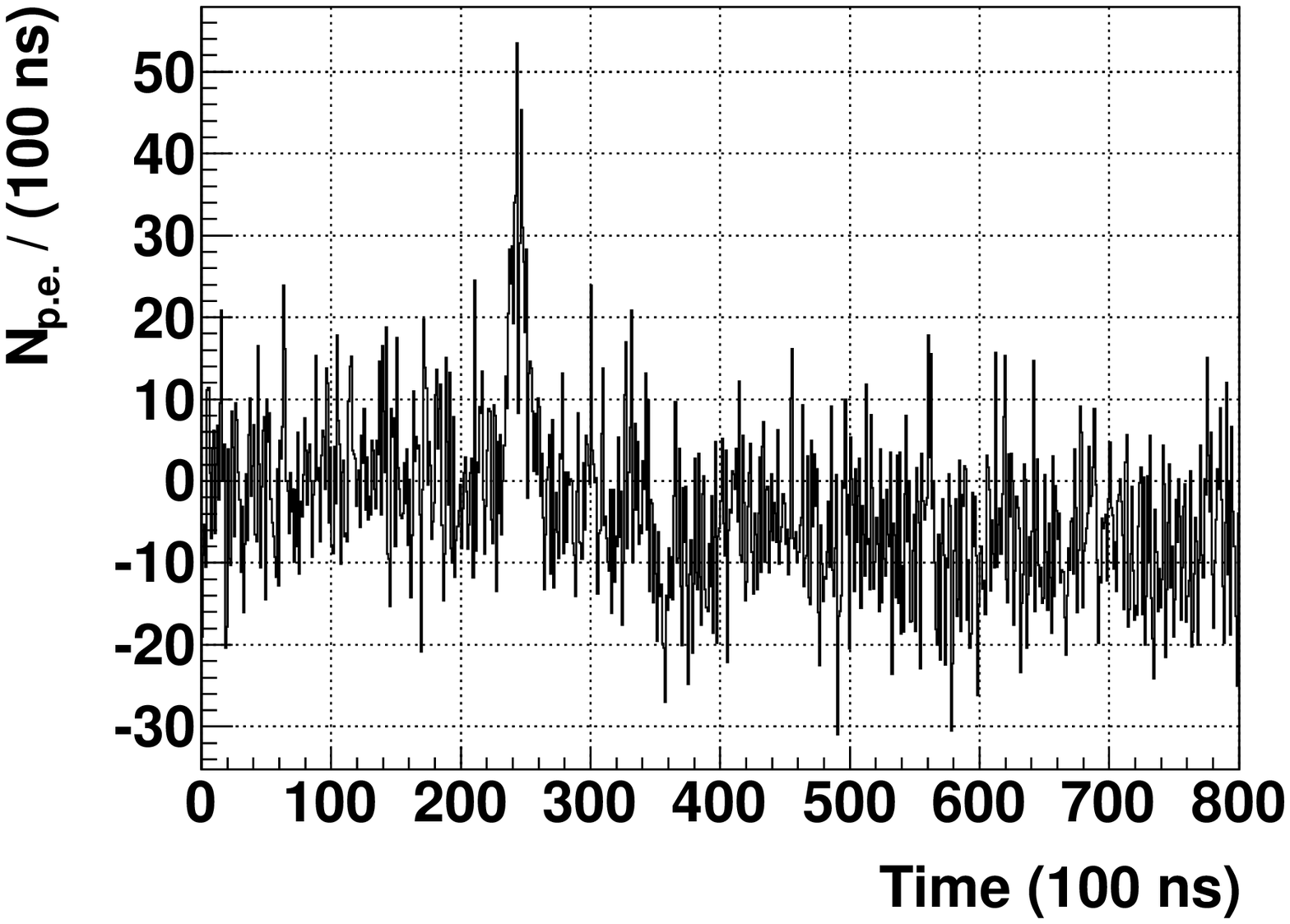}}
    \subfigure[Simulation: $E_{\rm{sim}}=10^{17.2}$ eV]{\includegraphics[width=0.49\linewidth]{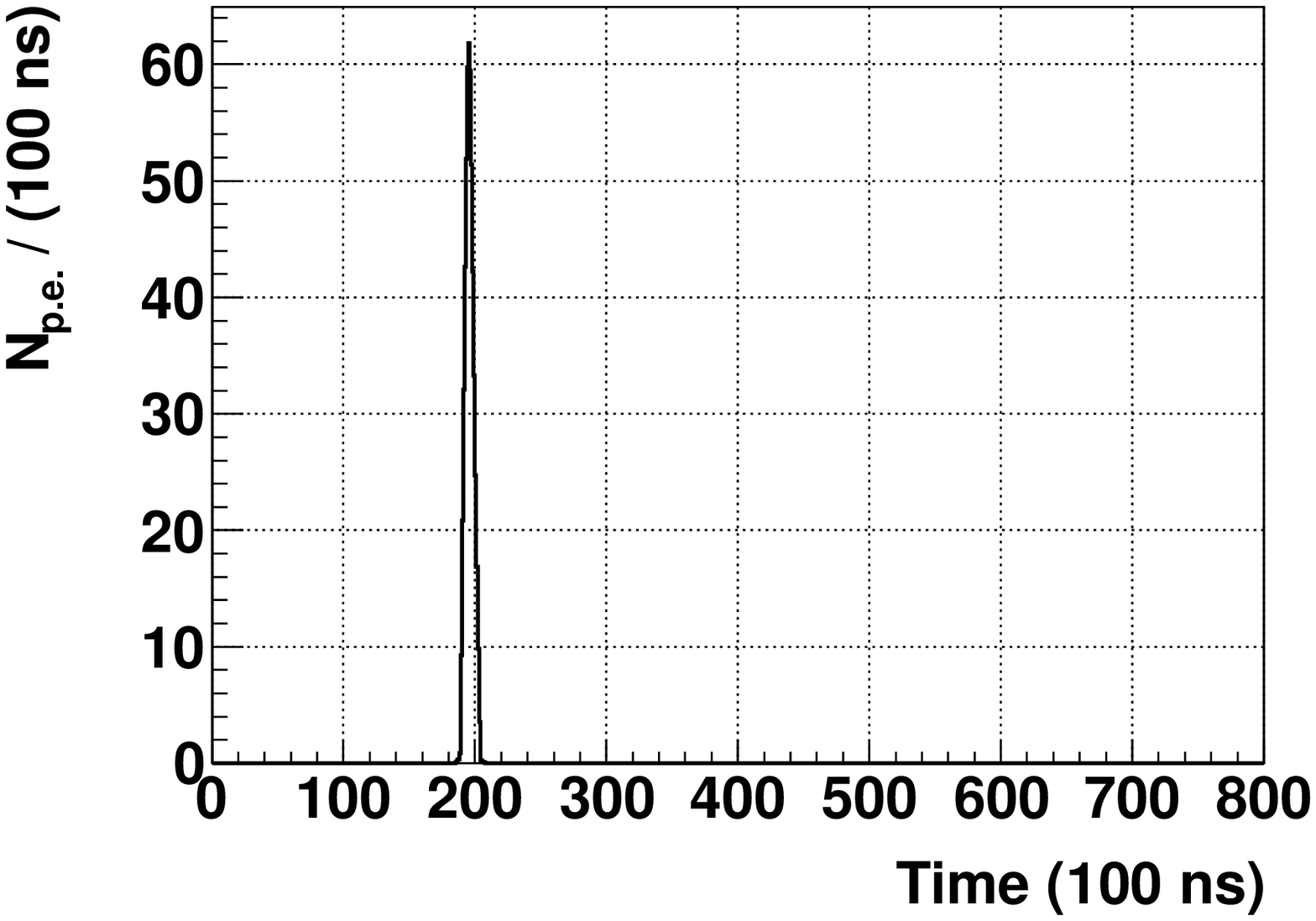}}
    \subfigure[Simulation: $E_{\rm{sim}}=10^{18.0}$ eV]{\includegraphics[width=0.49\linewidth]{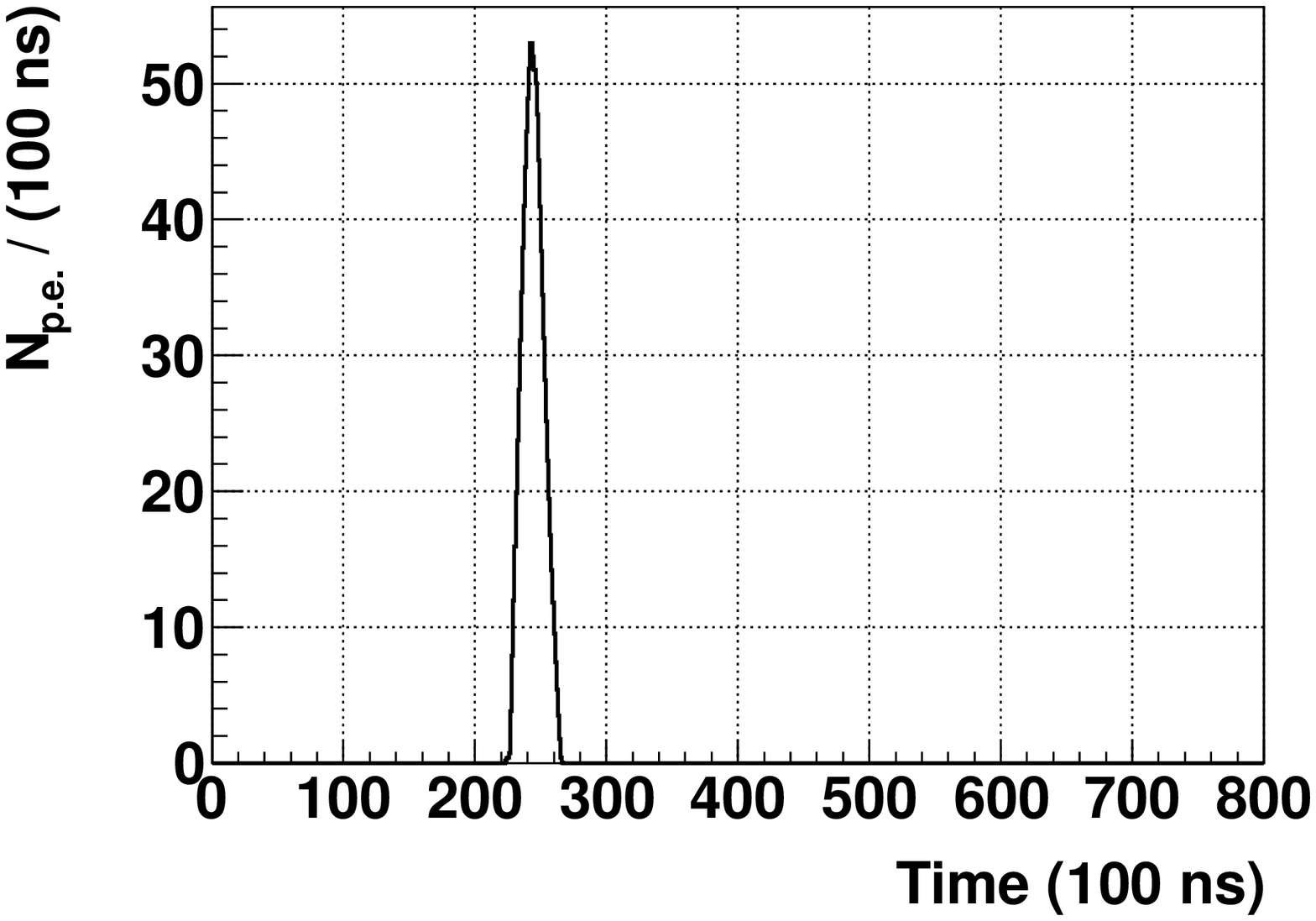}}
    \caption{(a) and (b): FADC signals recorded for two shower candidates; (c) and (d): corresponding simulated signals.}
    \label{fig:shower_datamc}
\end{figure}

Given the limited FOV of the FAST prototype, only a small portion of the shower development is actually observed and hence these low energy showers have their $X_{\rm{max}}$ located outside the FOV. Thus, a reliable Gaisser-Hillas fit to the shower profile is not possible. However, a comparison between the measured signal and simulations provides a useful cross-check. For each shower candidate we generated a shower with the same energy, direction and core position (as determined by the TA FD reconstruction). The corresponding FAST signal was simulated taking into account the telescope optics, the atmospheric attenuation and the PMT quantum efficiency as described in Section~\ref{subsec:laser}. Examples of simulated FAST traces  are given in Figure~\ref{fig:shower_datamc}, together with the measured traces of the corresponding candidate showers. The amplitude and shape of the simulated pulses are in good agreement with measurements.

\section{Conclusions and outlook}
\label{sec:conclusions}
We have presented a novel concept for an air shower fluorescence detector which features just a few pixels covering a large field of view. The FAST concept may be used in the next generation of UHECR experiments, which will require low-cost detectors to achieve an order of magnitude increase in aperture. Simulations indicate that UHECR showers with energies above $10^{19.5}$~eV will be detected by FAST with high efficiency and with resolutions comparable to current generation FDs. We have performed first tests of the FAST concept at the Telescope Array site where we installed a 200 mm PMT in the existing EUSO-TA telescope optics. The FAST prototype took data during 19 nights, for a total of 83 hours. The detector operated under a variety of conditions typical of field deployment (changes in temperature, night sky background and atmosphere; airplanes in the field of view; unexpected power cuts), demonstrating its stability and robustness. UV lasers placed at several kilometres distance were clearly detected by the FAST prototype, providing an estimate of its sensitivity. We also searched for UHECR showers detected by FAST in time-coincidence with the TA FD and found 16 highly significant candidates.  Simulations of the FAST response to lasers and UHECR showers show good agreement with the measurements, giving us confidence in the validity of the concept and its expected performance.

\begin{figure}
  \centering
    \includegraphics[width=0.5\linewidth]{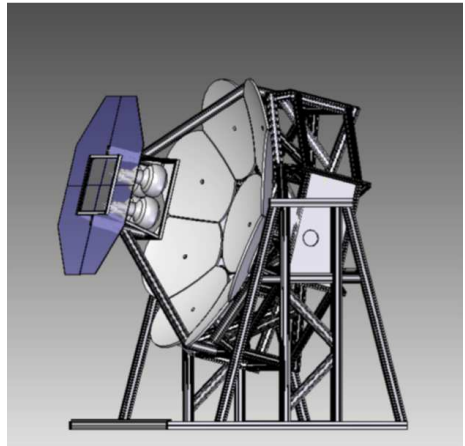}
    \caption{Preliminary design of a full-scale FAST prototype with a $30^\circ$ x $30^\circ$ FOV. }
    \label{fig:fast_design}
\end{figure}

Motivated by these encouraging results, a full-scale FAST prototype is under development. A preliminary design consisting of a $30^\circ \times 30^\circ$ FOV telescope of 1 m$^2$ effective area, with a 2$\times$2 PMT camera and a segmented spherical mirror of 1.6 m diameter is shown in Figure~\ref{fig:fast_design}.
We plan to install the full-scale FAST prototype at both the Auger and TA sites, where it will be able to provide measurements of showers that are largely independent of those made by the currently installed FDs (apart from the geometry supplied by the surface array), to cross check their energy and $X_{\max}$ measurements. Furthermore, FAST could be used as a useful cross check of the FD calibrations of these two experiments.

\section*{Acknowledgements}
This work was supported in part by NSF grant PHY-1412261 and by the Kavli
Institute for Cosmological Physics at the University of Chicago through
grant NSF PHY-1125897 and an endowment from the Kavli Foundation and its
founder Fred Kavli. This project has been partially funded by the Italian
Ministry of Foreign Affairs and International Cooperation. The Czech
authors gratefully acknowledge the support of the Ministry of Education,
Youth and Sports of the Czech Republic project No. LG13007. TF was
supported by the Japan Society for the Promotion of Science (JSPS)
Fellowship for Research Abroad H25-339.
The authors thank the Telescope Array and the JEM-EUSO Collaborations for
providing logistic support and part of the instrumentation to perform the
FAST prototype measurement. They also thank the Pierre Auger Collaboration
for fruitful discussions.

\bibliography{fast}

\clearpage 

\end{document}